\documentstyle[preprint,pre,aps,prl,epsfig]{revtex}
\tightenlines
\begin{document}     
\title{Switch of lamellae orientation in slits}
\author { A.Ciach and M.Tasinkevych} 
\address{Institute of Physical Chemistry and College of Science, 
Polish Academy of Sciences \\
Kasprzaka 44/52, 01-224 Warsaw, Poland \\} 
\date{\today} 
\maketitle 

\begin{abstract}
Effect of hydrophilicity of the confining walls on lamellar phases in 
oil-water-surfactant mixtures is studied in a slit geometry.
 In contrast to strongly hydrophilic or hydrophobic walls, which induce
 parallel orientation of lamellae, the lamellae can be oriented 
perpendicularly to the {\it neutral}
 walls when the material properties and the termodynamical state of the 
sample are suitably chosen. When  the elastic energy associated with 
compression or decompression of the lamellae parallel to very {\it weakly 
hydrophilic} walls   is sufficiently large,
 then  changes of the film thickness lead to  a switch from
 the parallel to the perpendicular orientation of the lamellae. Our general 
arguments are confirmed by explicit mean-field calculations in a lattice 
vector model.
\end{abstract}


In self-assembling systems such as diblock
copolymer melts, binary  or ternary surfactant mixtures, lipids etc., various
ordered structures occur spontaneously on a nanometer length scale 
\cite{matsen:96:0,laradji:97:0,gompper:94:0}.
The most common is the lamellar phase \cite{tiddy:80:0}.  
In ternary surfactant 
mixtures the lamellar
 phase  consists of oil- and water-rich layers separated by monolayers of
 surfactant. It is natural to expect that the layers are parallel to 
the confining walls and  the parallel orientation of lamellae is commonly
 assumed. Indeed, near a strongly hydrophilic or  hydrophobic 
wall a water-rich or an oil-rich layer is adsorbed at the surface 
respectively,
by which the orientation of the subsequent layers is fixed 
\cite{kekicheff:89:0,kekicheff:90:0,kekicheff:97:0,holyst:98:0}.
 However, when 
the surface is neutral, i.e. it is  neither hydrophilic nor hydrophobic, 
then 
neither component is favored and  no orientation of amphiphiles 
is preferred near the wall. Therefore  perpendicular rather than
parallel orientation of the lamellar phase may be stable between two parallel
 surfaces that are neither hydrophilic nor
hydrophobic (see Fig.1).  In the case 
of very weakly hydrophilic walls perpendicular
 orientation was indeed observed in Monte Carlo simulations \cite{holyst:98:0}
 of the phenomenological Landau-Ginzburg model \cite{gompper:94:0}.
Perpendicular orientation should be favoured  when the distance between the 
walls and the period of the lamellar phase do not match. On the other hand, 
 when the distance between the lamellae
 fits the distance between the weakly hydrophilic walls, 
parallel orientation  can be stable. 

 Films with the lamellae oriented  parallelly or 
 perpendicularly to the walls have different mechanical, 
electrical, optical etc. properties.  Therefore the possibility
 of inducing a switch between the two
 orientations by changing a control parameter such as the distance between
 the walls $L$  is potentially very important for various applications. 
For example, when some electrolyte is added, the electrical conductivity
 between the confining walls is high in the perpendicular orientation,
 whereas the parallel orientation resembles a sequence of capacitors.  
The switch of the lamellae orientation leads also to an abrupt change of the 
refraction index and the reflectivity, and of  elastic properties of the film.
A lamellar film containing a few lamellae  has the thickness
 $L\sim 30 \div 100 nm$. Such a
lamellar film  would 
correspond to a  switch of a  mesoscopic size -- a device which may be 
useful  for modern nanotechnologies. In this letter we show under what 
circumstances such a  switch from the parallel to the perpendicular 
 orientation of the  lamellae can occur.

The  confined fluid  induces the solvation force between the walls 
 \cite{evans:87:0},
\begin{equation}
\label{f}
f A=-\Bigg(\frac{\partial \Omega}{\partial L}\Bigg)_{\mu,T,A}-p A,
\end{equation}
where $\Omega, p, A,T$ and $\mu$ are the grand thermodynamic potential,
the  bulk pressure, the area of the confining walls, the temperature and
the  chemical 
potential respectively. $f$ can be directly measured in surface force apparatus
experiments
 \cite{kekicheff:90:0,kekicheff:97:0,israel:79:0,israel:81:0,abillon:90:0,petrov:94:0}. In simple fluids oscillating $f$ reflects packing effects of 
molecules for $L$ up to several molecular diameters, since the streched or 
shrunk confined fluid tends to assume the bulk structure. Similarly, but on a 
much larger length scale, $f$ reflects packing effects of oil- and
 water-rich domains in the complex fluids. For $L\ge 3\lambda$, 
where $\lambda$ is the period in the bulk phase,   lamellae are parallel to
the strongly hydrophilic (or hydrophobic) walls
\cite{kekicheff:90:0,kekicheff:97:0,tasinkevych:99:2}. 
 When $L$ and $\lambda$ do not match, the lamellar structure is stretched for
 $L>L_N$ or shrunk for  $L<L_N$ where 
$L_N=N\lambda+\lambda/2$ denotes the equilibrium
thickness,  corresponding to no stress, for a given number $N$ of periods of 
the confined phase. For $L\ne L_N$ the confined
 lamellar phase responds elastically to the stress strain, just like joined
strings,  each string representing one period of the lamellar phase 
\cite{kekicheff:97:0}.  The
  elastic contribution  to  the excess thermodynamic potential
\begin{equation}
\label{Oex}
\Omega^{ex}A=\Omega-\omega_b LA,
\end{equation}
where $\omega_b$ is the bulk thermodynamic-potential density,
has a form 
\begin{equation}
\label{Oel}
\Omega^{ex}_{el}= B\Delta L^2/2L_N,
 \end{equation}
with $ \Delta L=L-L_N$.  The
elastic modulus $B$ depends on the amphiphilicity of the surfactant and on 
the thermodynamic state \cite{tasinkevych:99:2}.
 
 Because neither component is favoured near the  neutral wall, 
it has a disordering effect on the lamellar order due to the missing neighbors
\cite{ciach:99:0}.  The missing-neighbors contribution to the excess
 potential $\Omega^{ex}$
 depends on the orientation of the lamellar phase (see Fig.1). 
For the parallel orientation ($\parallel$),
  $\Omega^{ex}=\Omega^{ex}_{\parallel}$, 
the interactions are cut along a single layer or 
along an interface. For the perpendicular orientation ($\perp$),
  $\Omega^{ex}=\Omega^{ex}_{\perp}$, the water-water,
 surfactant-surfactant and oil-oil
 interactions are missing at the surface in proportions depending on 
$\lambda$ and the thickness of the monolayer $a$. The density 
distributions near the surfaces tend to minimize the effect of the
 missing neighbors at the walls, and determine the values of the surface 
tensions $\sigma_{\parallel}$ and $\sigma_{\perp}$.
  Sufficiently far from phase transitions the effect
 of the wall should extend to distances comparable with the characteristic 
length of the system, which in this case corresponds to $\lambda$. 
If the lamellae are perpendicular 
to the confining walls, there is no constrain on the period of the lamellar
 phase, which can be equal to  the bulk value, and in this case there is no 
elastic contribution to  $\Omega^{ex}_{\perp}$.
Thus except for narrow
 slits, i.e. for  $L> 3\lambda$, we expect
 $\Omega^{ex}_{\perp}\approx 2\sigma_{\perp}$  independently of $L$. 
In the case of the parallel orientation of 
lamellae, for $L\ne L_N$, there are deformations of the structure in the whole
 slit  like in the case of the hydrophilic walls and
$\Omega^{ex}_{\parallel}\approx 2\sigma_{\parallel}+\Omega^{ex}_{el}(L)$.

 By suitable tuning of  hydrophilicity of surfaces, amphiphilicity, $T$ 
and surfactant volume-fraction $\rho_s$ one could in 
principle obtain comparable $\sigma_{\parallel}$ and 
$\sigma_{\perp}$. For weakly hydrophilic walls one can expect 
 $\sigma_{\parallel}< \sigma_{\perp}$. Thus, the parallel orientation should be
 stable in the case of no stress ($L=L_N$). 
The perpendicular  orientation may be
stable for slightly different wall separations,  corresponding to stretched 
or
shrunk parallel lamellar phases, if the elastic contribution to 
$\Omega^{ex}_{\parallel}$ is sufficiently large.
According to the phenomenological theory \cite{helfrich} 
$B=9\pi^2(kT)^2\lambda/2[64\kappa(\lambda/2-a)^4]^{-1}$, where $\kappa$ is the 
bending elasticity of monolayers.
 Thus the compression or decompression induced 
switch to the perpendicular orientation
is to be expected for relatively small  $\lambda$,
 rather than for highly swollen lamellar phases. 

Whether the switch can indeed
occur in particular systems should be verified either experimentally or by 
a more accurate analysis. The theoretical analysis should  allow for a 
determination of  $\sigma_{\perp},
\sigma_{\parallel}$ and $\Omega_{el}$ (or more accurately
$\Omega ^{ex}_{\parallel},\Omega ^{ex}_{\perp}$) in particular systems. 
There are several theoretical approaches to bulk self-asembling systems.
The approach suitable for a particular problem depends on the relevant
 length scale. In our case two length ratios are important, namely 
$L/\lambda$ and $\lambda/a$. $L/\lambda$ is the number of oil (water) layers 
in the slit for the $\parallel$ orientation, and  $\lambda/a$ characterizes the
liquid-crystalline lamellar phases. For $\lambda/a$ sufficiently large 
( $\lambda/a>4$), the water-rich 
layers are sufficiently thick to be compressible as in bulk liquid. 
Experimental results
\cite{kekicheff:90:0,kekicheff:97:0,israel:79:0,israel:81:0,abillon:90:0,petrov:94:0} show that the  lamellar phases
respond to compression or decompression by shrinking or swelling the 
water-rich (and/or oil-rich) layers,
 rather than by developing deformations characteristic for solids or by 
deformations of the molecular structure of amphiphiles. Hence the degrees of
 freedom related to the structure of
 molecules can be neglected, as long as $L/a\gg 2$. Dislocations, frustration
etc. can be expected for stiff, solid-like structures, i.e. when the
 water-rich  (oil-rich) layers are very thin, $\lambda/a< 4$. Since for our
 problem the relevant length rations are $1<L/\lambda\le 10$ and
 $4< \lambda/a\le 10$, the molecular degrees of freedom, frustrations etc.
 can be disregarded. On the other hand, the popular approaches in which the
 monolayers are approximated by infinitely thin mathematical surfaces may be 
oversimplified, particularly for  comparable thicknesses of the  layers of 
water and surfactant ($4\le \lambda/a \le 6$).

In a presence of a wall the surface term describing the interaction of the
 surface with the water, oil, and amphiphiles in various orientations must be
 included. In a purely phenomenological approach one cannot apriori know what
 values of  additional surface parameters are physical. Therefore one cannot
 know if the predicted phenomena, occurying for  certain values of parameters,
 are physical or not. It is thus advantegeous to consider a semi-microscopic
 approach and relate the interactions with the surface to the effective 
interactions in the bulk. For example, for the water-covered surface the 
interactions between various components and the surface are determined by
 the interactions between these components and water and there is much less 
ambiguity in defining the surface contribution in such an approach. 

All the above discussed requirements for the appropriate description of 
confined amphiphilic systems  are 
fullfiled by the semi-microscopic 
 CHS (Ciach, H\o ye and Stell) lattice model \cite{ciach:88:0}. The model
 is described in 
detail in Ref.\cite{ciach:88:0} and here we remind it only briefly. 
The microscopic states are
 $\hat{\rho}_{\it i}({\bf x})=1 (0) $ if the site
 ${\bf x}$ is (is not) occupied by the state $i$, where $i=1,2,...,2+M$ 
denotes water, oil and surfactant in different orientations respectively. 
In the case of close-packing and oil -- water symmetry only one chemical 
potential variable is relevant, namely 
$\mu=\mu_1-\mu_{surf}=\mu_2-\mu_{surf}$, with 
$\mu_{surf}=\mu_i$ for $i>2$.  The Hamiltonian in a presence of external
 fields $h_i({\bf x})$  can be written as:
\begin{eqnarray}
H=\frac{1}{2}\sum_{\bf x \neq x^{\prime}}\sum_{\it i,j}
\hat{\rho}_{\it i}(\bf x \rm ) \rm U_{\it ij}(\bf x-x^{\prime})
\rm \hat{\rho}_{\it j}(\bf x^{\prime} \rm )+ \nonumber \\ 
+\sum_{\bf x\rm}\sum_{\it i}h_{\it i}(\bf x\rm)\hat{\rho}_{\it i}(\bf x \rm )-
\mu \sum_{\bf x\rm}\Bigl (\hat{\rho}_1(\bf x \rm )+
                           \hat{\rho}_2(\bf x \rm )\Bigr ).
\label{H}
\end{eqnarray}
The lattice constant 
$a\equiv1$ is  identified with the length of the amphiphiles ($\sim 2 nm$).
 Nearest-neighbor interactions are assumed and 
$ \rm U_{\it ij}(\bf x-x^{\prime})$ vanishes for
 $|{\bf x}-{\bf x}'|\neq 1$.  In the case of oil-water symmetry
the water-water and oil-oil interactions are of the same strength 
 $-{\sf b}$, and the water-oil interaction energy is set to zero. The 
interaction between an amphiphile in an orientation $\hat \omega$ at 
 ${\bf x}$
 and a water (or oil) particle at  ${\bf x}'$ is  
$-{\sf c}\hat \omega \cdot \Delta {\bf x}$ (or $+{\sf c}\hat \omega \cdot \Delta {\bf x}$),
 where $ \Delta {\bf x}={\bf x}-{\bf x}'$, thus
 opposite orientations
 of amphiphiles are preferred by the water and the oil particles, as in real 
systems. Two amphiphiles with orientations  $\hat \omega$ and  $\hat \omega'$
 at  ${\bf x}$ and  ${\bf x}'$ respectively contribute
 $-{\sf g}(\hat \omega\times \Delta{\bf x})
 \cdot(\hat \omega'\times \Delta{\bf x})$
 to the system energy (when
 $| \Delta{\bf x}|=1$), i.e.  ${\sf g}$ supports formation of planar 
monolayers with amphiphiles parallel to each other and perpendicular to the 
surface they occupy.  Explicit expressions for 
$ \rm U_{\it ij}(\bf x-x^{\prime})$ defined above can be found in
 Ref.\cite{tasinkevych:99:2}.
For lamellar phases the direction perpendicular to the planar layers of 
surfactant, oil and water is distinguished. We arbitrarily choose the 
orientation of the unit vector normal to the layers and denote it by 
$\hat {\bf n}$. Then all orientations of amphiphiles such that 
$\hat \omega\cdot\hat {\bf n}>0$ (or $\hat \omega\cdot\hat {\bf n}<0$) 
are mapped onto a single state
 $\rightarrow$ (or $\leftarrow$). Hence, once the direction of density 
oscillations is fixed,  there are 4 states corresponding to water,
 oil,  $\rightarrow$ and  $\leftarrow$  at every lattice site.

The model parameters characterize the effective interactions between 
different components, and the interactions with the surface can be expressed
 in terms of these parameters. When the surface is water-covered, all the 
components interact with the wall in the same way as with the bulk water, and 
$h_i({\bf x})$ can be easily derived from the bulk interactions.
 When the hydrophilicity of the wall is decreased, the interactions with the 
wall can be  decreased by the same factor $h_s$ for all the components.
 Thus, $0\leq h_s\leq 1$ should correspond, at least qualitatively,
 to physical surfaces, ranging from
 neutral through weakly hydrophilic to strongly hydrophilic walls. 

The model parameters are not directly measurable, but the calculated values
 for measurable quantities can be compared with experiments. For example, 
the model parameters can be choosen so that 
$\lambda/a$ is the same as in a particular experimental system. In this way
 the model parameters can be related to different experimental systems in the 
bulk. It turns out that once the
 bulk properties in the model and experiments are the same, for strongly 
hydrophilic walls (the case with no free surface parameters) the modulus of 
compressibility $B$ (in $kT$ units) in the model and experiments agree very 
well \cite{tasinkevych:99:2}.
Therefore, in contrast to purely phenomenological models with arbitrary 
parameters, the CHS model can quite reliably  predict phenomena not yet 
studied experimentally.

The structure of the stable confined lamellar phase in 
mean-field approximation (MF)
 corresponds to the global minimum of  the grand-thermodynamic potential
\begin{eqnarray}
\Omega^{MF}(T,\mu,\it L)=
\sum_{\bf x}\sum_{\it i}\rho_{\it i}({\bf x} )
\Biggl (kT \ln(\rho_{\it i}({\bf x }))+\rm \frac{1}{2}\phi_{\it i}({\bf x})+
\it 
h_{\it i}({\bf x})-\mu(\delta_{\it i\rm 1}+\delta_{\it i\rm 2})\Biggr ).
\label{W}
\end{eqnarray}
Here  
 $\phi_{\it i}({\bf x})=\sum_{\bf x^{\prime}}\sum_{\it j}\rm U_{\it ij}
 (\bf x-x^{\prime}\rm )\rho_{\it j}(\bf x^{\prime} \rm)$  is the mean field 
 and $\rho_i({\bf x})$ is the MF-average of $ \hat \rho_i({\bf x})$.
In the slit geometry we assume $1\le z\equiv x_3\le L$.
 We consider two classes of density distributions,
 corresponding to the parallel and the perpendicular orientations of the
 lamellar phase. In the first case the two distinguished orientations of 
amphiphiles are 
$\hat \omega =\pm \hat z$.
In the second case we assume that
 the layers are parallel to the $(y,z)$-plane, and
 $\hat \omega = \pm \hat x$.
  The slit 
is infinite in directions $x\equiv x_1$ and 
$y\equiv x_2$. In the  case of the parallel orientation of the lamellar phase
 $\rho_i({\bf x})\equiv \rho_i(z)$. 
In the second case, due to the deformations of the
 structure near the surfaces, the density distributions are 
$\rho_i({\bf x})\equiv \rho_i(x,z)$.
 In the direction $x$ we assume periodic boundary
 conditions in a system of a size $\lambda$, since $\Omega^{ex}$ for such 
system is the same as for infinite periodic structure with the period
 $\lambda$ (we also considered smaller 
and larger sizes to be sure we find the stable structure). 
 We use the method of finding local
 minima of $\Omega^{MF}$  by solving numerically a set of self-consistent 
 equations for densities,   tested in earlier works  
\cite{tasinkevych:99:2}. 
Then  we find the structures corresponding to the lowest
 values of   $\Omega^{ex}_{\parallel}$ and 
 $\Omega^{ex}_{\perp}$, and eventually we calculate 
 $\Omega^{ex}_{\parallel}-\Omega^{ex}_{\perp}$ to find the global minimum. 
It turns out that 
 $\Omega^{ex}_{\parallel}-\Omega^{ex}_{\perp}$ depends sensitively on the
 hydrophilicity of the surfaces $h_s$, on the ratio of water-surfactant
 ${\sf c}/{\sf b}$ and surfactant-surfactant ${\sf g}/{\sf b}$ interactions
 and on the period of 
the lamellar phase, which for given 
interactions depends on $ \mu$ (or $\rho_s$) and $T$.
 We find that for highly swollen lamellar 
phases the parallel orientation is preferred, and that
 for smaller periods $\lambda$   the perpendicular 
orientation stabilizes between neutral walls. We find that there exist
 systems for which the switch
 of the orientation does take place. Fig.2 refers to such a system and 
corresponds to the stability region of the lamellar phase far from phase 
boundaries, with $\lambda/a=6$.
The switch takes place for relatively small $\lambda$ and for weakly 
hydrophilic walls, as expected.
 Note that in the case of the neutral walls the period of
 $\Omega^{ex}_{\parallel}(L)$ is $\lambda/2$ (Fig.2a),
 since either oil- or water-rich layers 
can be adsorbed at the 
surfaces. In contrast, in the case of the hydrophilic walls only the
 water-rich layers are formed near the walls and the period of 
$\Omega^{ex}_{\parallel}(L)$ is  $\lambda$ (see Fig.2c). 

In MF the 
correlations between the lamellae are neglected. Such correlations are 
particularly important close to the melting of the lamellar phases
 where they lead to creation of channels connecting the neighboring water-rich
 layers and eventually to the transition to the microemulsion.
 Far from phase boundaries, however, the correlations should play less 
important role. In the case of the hydrophilic walls the MF results for the 
CHS model agree quite well with the results of experiments. The effect of the 
correlations should not depend significantly on the kind of the confining 
walls and we expect that the correlations can have quantitative, but not 
qualitative effect on the results. 

Our results show that the switch is to be expected for $\lambda/a\approx 6$, 
i.e. with the water (oil) layers twice as thick as the surfactant monolayers,
 and sufficiently far from phase boundaries. From the rough relation 
$2a/\lambda\approx \rho_s$ we obtain the estimation 
$\rho_s\approx 0.33\pm 0.1$ for the surfactant volume
 fraction  which should correspond to the occurence of the switch. For this 
range of $\rho_s$ the lamellar phase is stable at room temperatures for 
example for extensively studied experimental system such as water, decane and
$C_{10}E_{5}$ \cite{kahlweit:86:0}.

Various experimental  methods can be used for detecting the switch,
 due to the abrupt
 change of various propertis of the sample. For example, SFA \cite{israel:79:0}
 measurements should show vanishing force for the perpendicular orientation
 and elastic responce for the parallel orientation. Reflectivity, index of 
refraction and electrical conductivity change abruptly when the switch takes 
place.  Measurememts of the above quantities should give significantly 
different results for the two orientations, thus allowing for observation of
 the phenomenon.

{\bf Acknowledgment}
We would like to thank Prof.R.Ho\l yst for discussions and comments.
This work was partially supported by the KBN grant 3T09A 073 16.

\newpage
\centerline{\epsfig{file=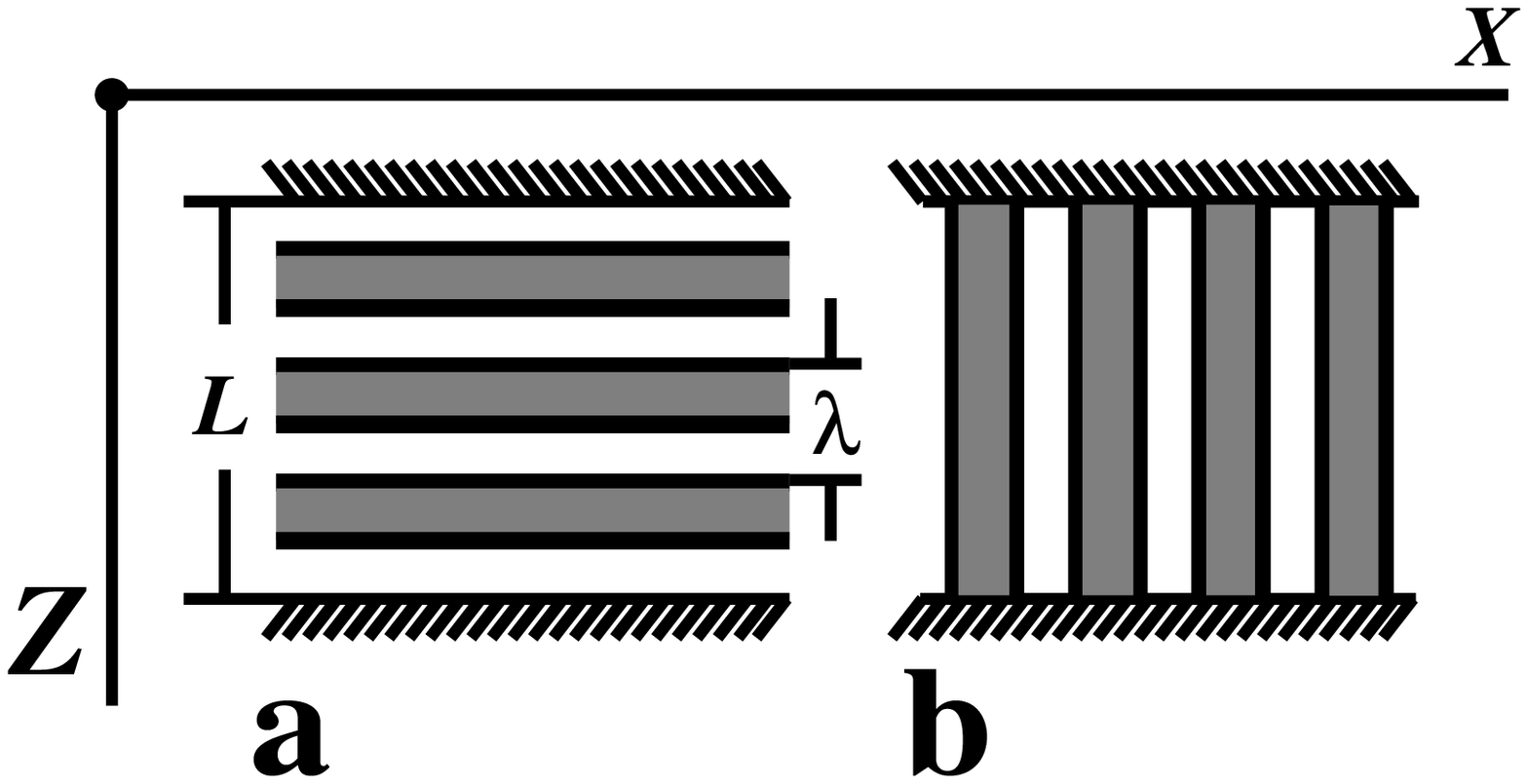,width=15cm,angle=0}}
\vspace*{1cm}

{\bf Fig.1.}
 {Schematic illustration of parallel (a) and perpendicular (b) orientations of the lamellar 
phase 
in a slit. Shaded regions represent oil-rich domains, while white regions  represent water-rich 
layers and thick black lines represent the surfactant monolayers.}

\newpage
\centerline{\epsfig{file=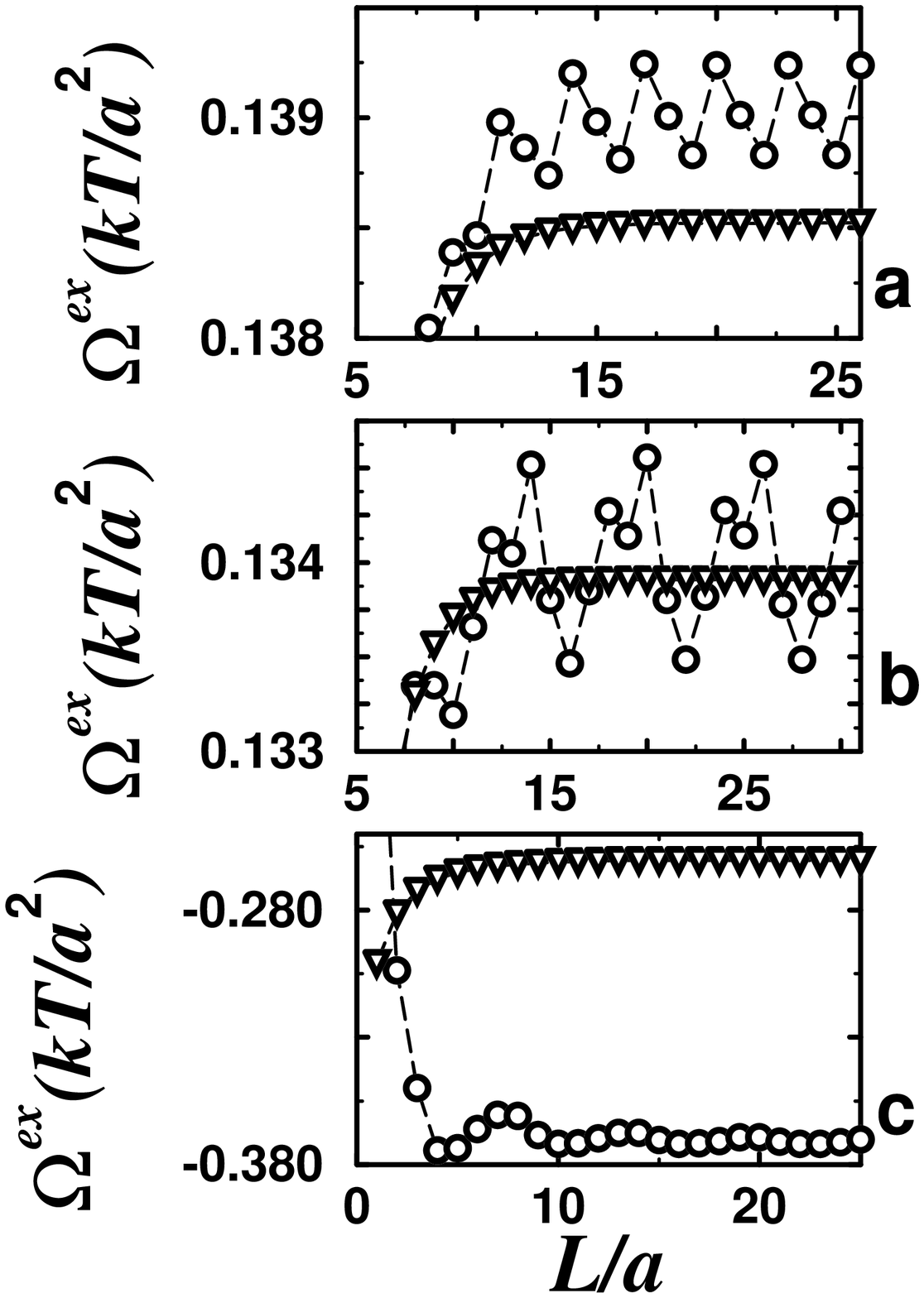,width=14cm,angle=0}}
\vspace*{1cm}

{\bf Fig.2.}
  {$\Omega^{ex}_{\parallel}(L)$ (white circles)  and  $\Omega^{ex}_{\perp}(L)$ (triangles down)
as a function of the wall separation (in units of $a$) for $\sf c/b\rm =2.4$, $kT/\sf b=\rm 2.8$, $\mu/\sf b=\rm 3$ and $\sf g/b\rm =0.15$.
 (a) $h_s=0$ (neutral walls);
(b) $h_s=0.015$ (weakly hydrophilic walls); (c) $h_s=1$ (strongly hydrophilic walls). }

\end{document}